\documentclass[floatfix,showpacs,epsfig]{revtex4}
\usepackage{pdfpages}
\usepackage{epsf}
\usepackage{subfigure}
\usepackage{amsmath}
\usepackage{epstopdf}
\usepackage{mathrsfs}
\usepackage{natbib}
\usepackage{amssymb,latexsym}
\usepackage{dcolumn}
\usepackage{graphicx}

\newcommand{\be}{\begin{equation}}
\newcommand{\ee}{\end{equation}}
\newcommand{\bea}{\begin{eqnarray}}
\newcommand{\eea}{\end{eqnarray}}
\begin{document}
\title{\large \bf Relativistic rotation curve for cosmological structures}

\author{Mohammadhosein Razbin}
\affiliation{ Georg-August-Universit\"{a}t G\"{o}ttingen, Institut f\"{u}r Theoretische Physik, Friedrich-Hund-Platz 1, 37077 G\"{o}ttingen, Germany, EU and\\
Max-Planck-Institut f\"{u}r Dynamik und Selbstorganisation, Am Fassberg 17, 37077 G\"{o}ttingen, Germany, EU }
\email{m.razbin@theorie.physik.uni-goettingen.de }

 \author{Javad T. Firouzjaee}
\affiliation{ School of Physics and School of Astronomy, Institute for Research in Fundamental Sciences (IPM), Tehran, Iran }
 \email{j.taghizadeh.f@ipm.ir}
\author{Reza Mansouri}
\affiliation{Department of Physics, Sharif University of Technology,
Tehran, Iran and \\
  School of Astronomy, Institute for Research in Fundamental Sciences (IPM), Tehran, Iran}
 \email{mansouri@ipm.ir}

\date{\today}

\begin{abstract}
 Using a general relativistic exact model for spherical structures in a cosmological background, we have put forward an algorithm to calculate the test particle geodesics
within such cosmological structures in order to obtain the velocity profile of stars or galaxies. The rotation curve thus obtained is based on a density profile
and is independent of any mass definition which is not unique in general relativity. It is then shown that this general relativistic rotation curves for a toy model
and a NFW density profile are almost identical to the corresponding Newtonian one, although the general relativistic masses may be quite different.
\end{abstract}
%
%
\maketitle
\section{Introduction }

There has been recently many attempts to understand fully the consistency of the linearized Einstein equations; in other
words how far we are right in using Newtonian versus general relativistic gravity in astrophysical and cosmological
applications (see for example \cite{relativity-newtonian} and \cite{wald}). We face this controversy in this paper by calculating
the general relativistic rotation curve for a spherical structure within a FRW universe. The exact solution gives us
the possibility to see if and in which sense the Newtonian approximation is valid in the case of the weak gravity. \\
The relativistic rotation curve has been the subject of some recent papers (see \cite{Cooperstock1} and \cite{Cooperstock2}).
The model discussed in these papers, however, is based on the simple Schwarzschild metric, dust collapse, or a
simple axially symmetric space-time with a singularity at the central plane, not taking into account the full
dynamics of a cosmological structure. We, however, look for an exact solution of Einstein equations representing
a cosmological structure as an overdensity region within a cosmological background assumed to be asymptotically FRW.
There are not much viable exact models representing such a structure. We will use an analytical
model proposed recently based on an inhomogeneous cosmological LTB solution \cite{javad1}. Its time-like
geodesics are then integrated numerically to obtain the rotation curve of the cosmological structure. Obviously
we need a specified density profile for our cosmological structure to obtain the rotation curve, without a need of specifying a mass. In the Newtonian case
this is identical to have a unique mass of the structure. In general relativity, however, there is no
unique mass definition for astrophysical objects given a specific density profile\cite{javad2}. Therefore, relying on a density profile which is an observational quantity, the
non-uniqueness of the concept of mass does not disturb our conclusion about the comparison of the Newtonian versus
general relativistic rotation curve.\\

Section II is a brief introduction to the LTB metric, followed
by definitions of some quasi-local masses. In section III we introduce an algorithm how to calculate the rotation curve for a test particle
motion around a cosmological structure within a FRW universe. Section IV is devoted to the calculation of the general relativistic rotation 
curve for a cosmological structure using this algorithm and to compare the results with the Newtonian case. Some general relativistic quasi-local 
masses are also calculated and compared to the corresponding Newtonian one for the same density profile. We then conclude in section V. Throughout 
the paper we assume $8 \pi G = c = 1$.

\section{LTB model of a structure}

The LTB metric in synchronous coordinates is written as
\begin{equation}
 ds^{2}=-dt^{2}+\frac{R'^{2}}{1+f(r)}dr^{2}+R(r,t)^{2}d\Omega^{2},
\end{equation}

representing a pressure-less perfect fluid satisfying
\begin{eqnarray}\label{ltbe00}
\rho(r,t)=\frac{2M'(r)}{ R^{2}
R'},\hspace{.8cm}\dot{R}^{2}=f+\frac{2M}{R}.
\end{eqnarray}
Here dot and prime denote partial derivatives with respect to the
parameters $t$ and $r$ respectively. The angular distance $R$,
depending on the value of $f$, is given by
\begin{eqnarray}\label{ltbe1}
R=-\frac{M}{f}(1-\cos \eta(r,t)),\nonumber\\
\hspace{.8cm}\eta-\sin \eta=\frac{(-f)^{3/2}}{M}(t-t_{b}(r)),
\end{eqnarray}

for $f < 0$,
 and
\begin{equation}\label{ltbe2}
R=(\frac{9}{2}M)^{\frac{1}{3}}(t-t_{b})^{\frac{2}{3}},
\end{equation}
 for $f = 0$, and
\begin{eqnarray} \label{ltbe3}
R=\frac{M}{f}(\cosh \eta(r,t)-1),\nonumber\\
\hspace{.8cm}\sinh \eta-\eta=\frac{f^{3/2}}{M}(t-t_{b}(r)),
\end{eqnarray}
for $f > 0$.\\
The metric is covariant under the rescaling
$r\rightarrow\tilde{r}(r)$. Therefore, one can fix one of the three
free functions of the metric, i.e. $t_{b}(r)$, $f(r)$, or $M(r)$.
The function $M(r)$ corresponds to the Misner-Sharp mass in general
relativity ({\cite{mis-sha}, see also \cite{javad2}). The $r$ dependence
of the bang time $t_{b} (r)$ corresponds to a non-simultaneous big-bang or big-crunch singularity. \\
There are two generic singularities of this metric, where the
Kretschmann and Ricci scalars become infinite: the shell focusing
singularity at $R(t,r)=0$, and the shell crossing one at
$R'(t,r)=0$. However, there may occur that in the case of $R(t,r)=0$
the density $\rho = \frac{M'}{R^{2}R'}$ and the term $\frac{M}{R^3}$
remain finite. In this case the Kretschmann scalar remains finite
and there is no shell focusing singularity. Similarly, if in the
case of vanishing $R'$ the term $\frac{M'}{R'}$ is finite, then the
density remains finite and there is no shell crossing singularity either (see \cite{javad1} for more detail).\\
For our model structure, depending on our model calculation, we arrive in the case of a toy model at a specific
density profile, or specify a density profile like the NFW one from the beginning. In each case there is a unique
Newtonian mass. However, in the relativistic case we may allocate different masses without any preference. Although this
does not change our aim of a comparison between the relativistic rotation curve and the Newtonian one, we prefer to show how
different these masses may be. Note that the observational quantity is the density profile and not the total mass which
is a derived quantity. In fact more than 10 general relativistic mass definitions have already been proposed in the
literature. The difference between some of the proposed quasi-local mass definitions has been studied
in \cite{javad2}, where it has been shown that in the case of spherically symmetric structures the Hawking quasi-local mass is equal to
the Misner-sharp one. The Brown-York mass for the 2-boundary specified by $r=constant$ and $t=constant$ ($M_{BY.r}$) in an
asymptotically FRW solution is given by
\\
\begin{eqnarray} \label{ltbe3}
M_{BY.r}=-R\sqrt {1+f}+(R\sqrt{1+f})\mid _{FRW}.
\end{eqnarray}\\
If we specify the 2-boundary by $R(r,t)=costant$ and $t=constant$ with $R$ being the areal radius, then
the Brown-York mass ($M_{BY.R}$) is given by
\\
\begin{eqnarray} \label{ltbe3}
M_{BY.R}=-R\sqrt {1+\frac{2M}{R}}+(R\sqrt{1+\frac{2M}{R}})\mid _{FRW}.
\end{eqnarray}\\
The Liu-Ya and Epp masses are equal in our spherically symmetric case and are given by
\\
\begin{eqnarray} \label{ltbe3}
M_{LY}=M_{Epp}=-R\sqrt {1+\frac{2M}{R}}+(R\sqrt{1+\frac{2M}{R}})\mid _{FRW}.
\end{eqnarray}\\

We therefore concentrate on three mass definitions: the Misner-Sharp one which is equal to that of Hawking;
The Epp mass being equal to Liu-Yau and the Brown-York mass at either the constant areal radius ($R=constant$) or
the constant comoving radius ($r=constant$).

\section{Constructing the model}

The model we are going to construct and study should describe a simple model of a spherically symmetric mass condensation within
a FRW universe as an exact solution of Einstein equations with a pressure-less ideal fluid. Therefore we will choose a density profile
reflecting an overdensity at the center and almost constant density far from the center as expected for a FRW universe. Within such a
model structure we then study timelike geodesics to extract information about the rotation curve in such a dynamical setting. This is
achieved by specifying the three LTB functions $t_{b}(r)$, $f(r)$, and $M(r)$. Assuming an expanding universe, this cosmic LTB model structure
starts expanding with the universe before its expansion decouple from the universe and a collapsing phase starts. There are different
ways to specify the LTB functions depending on our needs or our methodology. Once the solution is fixed we may study the timelike
geodesics to extract the rotation curve within the structure. Now, the geodesic equations for the LTB metric are given by

\begin{equation}
\frac{d^{2}r}{d{\lambda}^{2}}=\frac{R(1+f)}{R'}[\sin(\theta)^{2}({\frac{d\phi}{d{\lambda}}})^{2}+({\frac{d\theta}{d{\lambda}}})^{2}]-\frac{[R'R''-\frac{{R'}^{2}f'}{2(1+f)}]}{{R'}^{2}}{({\frac{dr}{d{\lambda}}})^{2}}-2\frac{\dot{R'}}{R'}{({\frac{dt}{d{\lambda}}})({\frac{dr}{d{\lambda}}})},
\end{equation}
\begin{equation}
{\frac{d^{2}\theta}{d{\lambda}^{2}}=\sin(\theta)\cos(\theta)({\frac{d\phi}{d{\lambda}}})^{2}-2\frac{R'}{R}({\frac{dr}{d{\lambda}}})({\frac{d\theta}{d{\lambda}}})-2\frac{R'}{R}({\frac{dt}{d{\lambda}}})({\frac{d\theta}{d{\lambda}}})},
\end{equation}
\begin{equation}
\frac{d^{2}\phi}{d{\lambda}^{2}}=-2\frac{\cos(\theta)}{\sin(\theta)}(\frac{d\theta}{d{\lambda}})(\frac{d\phi}{d{\lambda}})-2\frac{R'}{R}(\frac{dr}{d{\lambda}})(\frac{d\phi}{d{\lambda}})-2\frac{R'}{R}(\frac{dt}{d{\lambda}})(\frac{d\phi}{d{\lambda}})
\end{equation}
and
\begin{equation}
\frac{d^{2}t}{d{\lambda}^{2}}=-{R{\dot{R}}{\sin(\theta)}^{2}({\frac{d\phi}{d{\lambda}}})^{2}} -R\dot{R}({\frac{d\theta}{d{\lambda}}})^{2}-\frac{R'\dot{R'}}{1+f}({\frac{dr}{d{\lambda}}})^{2}.
\end{equation}

These equations can be simplified by choosing the particle rotation plane in the $\theta=\pi/2$ and by the assumption of the geodesic being time like:
 $$ -1= -(\frac{dt}{d\lambda})^2+ \frac{R'^2}{1+f} (\frac{dr}{d\lambda})^2+ R^2(\frac{d\theta}{d\lambda})^2+R^2  \sin^2(\theta)(\frac{d\phi}{d\lambda})^2. $$
  Using these geodesic equations, we are able to find dynamical properties of a test particle within this structure.
 Given that our cosmic structure is dynamic we choose the initial conditions such that the structure is in its late
 dynamic phase, where the time scale of the evolution of the structure is greater than the time of revolution of the circular
 path of a test particle. Therefore, we may expect to have quasi-circular geodesics. These are defined either as those
 having a vanishing radial velocity and acceleration respect to \textit{areal radius} at the initial conditions, or have an almost constant radius
 within the numerical precision for a finite angular displacement. We have checked both procedures leading to
 the same numerical result. Although we have calculated the particle path, its velocity and the rotation curve at
 distances far from the center using general relativity, it is obvious that these local results should be well
 within the Newtonian approximation. Now, using the following Newtonian relation, we may define the Newtonian mass
 corresponding to the structure leading to the circular velocity reflected in the geodesic equations:
 \begin{equation}
M_N(R)={R}^{2}[{R}{(\frac{d\phi}{d\lambda})}^{2}-\frac{{d}^{2}R}{d{\lambda}^2}].
\end{equation}
where $R$ is the areal radius with $\lambda$ its affine parameter, and the acceleration given by
\begin{equation}
\frac{{d}^{2}R}{d{\lambda}^2}=R''(\frac{dr}{d\lambda})^{2}+2{\dot{R'}}(\frac{dr}{d\lambda})(\frac{dt}{d\lambda})+R'({\frac{d^{2}r}{d{\lambda}^{2}}})+{\ddot{R}}(\frac{dt}{d\lambda})^{2}+\dot{R}(\frac{{d}^{2}t}{d{\lambda}^{2}})
\end{equation}
The terms ${\frac{dt}{d\lambda}},\frac{dr}{d\lambda},\frac{d\phi}{d\lambda} $ may be calculated from the geodesic equations.\\
 In addition, we may use any general relativistic mass definition to calculate the relativistic quasi-local mass of the structure.
 Understanding the difference between these masses and the Newtonian one allocated to the structure and its rotation curve
 will be the subject of our discussion.

\section{Newtonian versus general relativistic rotation curve}

We report the results in three steps. To justify our algorithm and the code, we first apply it to the vacuum Schwarzschild case. We then go over
to a toy model structure, and finally apply the algorithm to a more realistic case with a NFW density profile.\\

\subsection{Rotation curve in the vacuum case of Schwarzschild}
Noting that the Schwarzschild space-time is a particular case of LTB space-time, we choose the three LTB functions in the following way:
\begin{equation}
f(r)=0,
\end{equation}
\begin{equation}
M(r)=m,
\end{equation}
\begin{equation}
t_{b}(r)=r.
\end{equation}
The resulting Schwarzschild metric is then given by
\begin{equation}
ds^{2}=-dt^{2}+\frac{1}{(\frac{3}{4}m(r-t))^\frac{2}{3}}dr^2+(2m)^\frac{2}{3}(\frac{3}{2}(r-t))^\frac{4}{3}({d\theta}^2+{sin(\theta)}^2{d\phi}^2 ).
\end{equation}\\
This is the metric of the Schwarzschild space-time in a synchronous frame. By studying circular geodesics having vanishing radial acceleration
and velocity, we obtain the rotation curve as given in the Fig.(\ref{fig1}). Here the mass is the unique Schwarzschild one, which is the same
as the Misner-Sharp mass. It is well known that in the case of Schwarzschild space-time at distances far from the center, i.e. $\frac{2M}{R}<<1$, the
Misner Sharp mass is equal to all the other general relativistic masses defined so far. Now, to obtain the Newtonian rotation curve we take the
Newtonian mass to be equal to the Schwarzschild one and then obtain the corresponding rotation curve. It is seen from Fig.(\ref{fig1}) that the
relativistic and the Newtonian rotation curves coincide as expected.

\begin{figure}[h]
\centering
\includegraphics[width=2.5in]{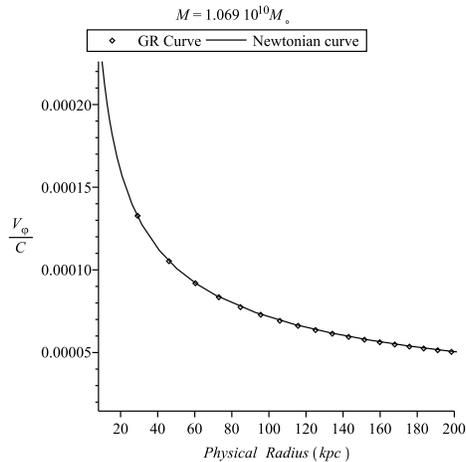}
\caption{General relativistic and the Newtonian rotation curves for the Schwarzschild metric } \label{fig1}
\end{figure}

\subsection{ A toy model}

Now we go over to study the rotation curve for a non-trivial but simple toy model. It is constructed as an exact
solution of the Einstein equations to represent a cosmic structure within a FRW universe. The three LTB functions are fixed in
 the following way\cite{javad1}:
\begin{eqnarray}\label{ltbe00}
M(r) =\frac {{r}^{3/2} \left( 1+{r}^{3/2} \right) }{a}, \hspace{.8cm}\ t_{b}(r)=0, \hspace{.8cm}\
 f(r)=-(\frac {r{{\rm e}^{-r}}}{b}).
\end{eqnarray}
The model represents a typical galaxy as a cosmological structure showing the formation of a central black hole
with distinct event- and apparent horizon. As expected for a cosmological structure, the density profile
shows a void before entering the asymptotic FRW region. Fig.(\ref{fig2}) shows the density profile of the
toy model and its corresponding relativistic and Newtonian rotation curve. The Newtonian rotation curve is
calculated using the mass related to the density profile at a specific radius. As we see from the figure, both
rotation curves are almost identical.\\
We note by passing that the density profile is obtained without fixing the mass which is not uniquely defined in
general relativity in contrast to the Newtonian case. Therefore, as our algorithm shows, the rotation curve may
be uniquely defined once we have fixed the density profile for a specific solution of Einstein equations. In other
words, any rotation curve within a cosmological structure corresponds to a unique density profile but not a unique
mass, in contrast to the Newtonian case. This is shown in Fig. (\ref{fig3}).

\begin{figure}[h]
\centering
\mbox{\subfigure{\includegraphics[width=2in]{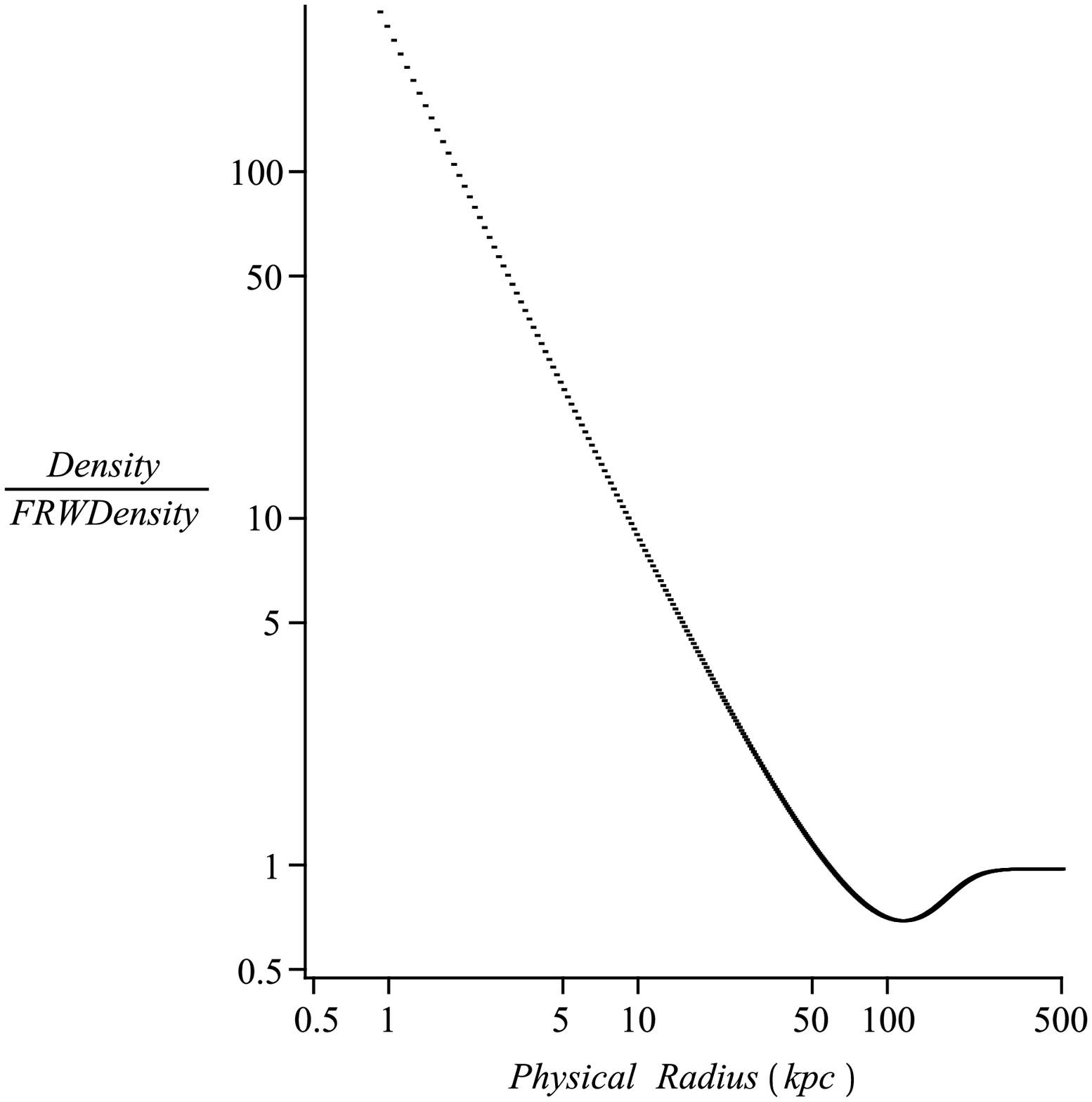}}\quad
\subfigure{\includegraphics[width=2in]{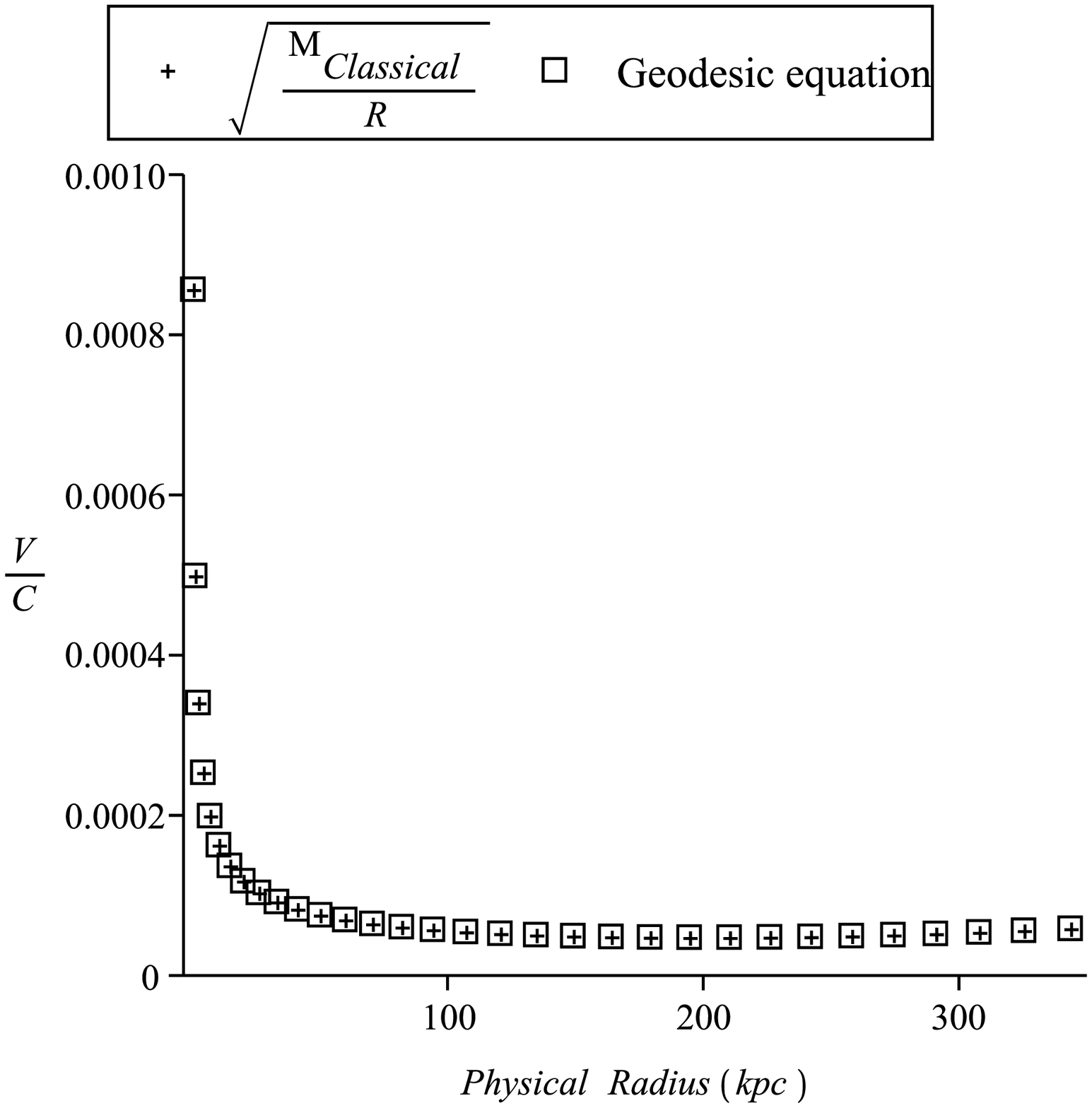} }}
\caption{ Density profile for the toy model of a typical galaxy and the corresponding general relativistic and Newtonian rotation curve} \label{fig2}
\end{figure}

\begin{figure}[h]
\centering
\mbox{\subfigure{\includegraphics[width=2in]{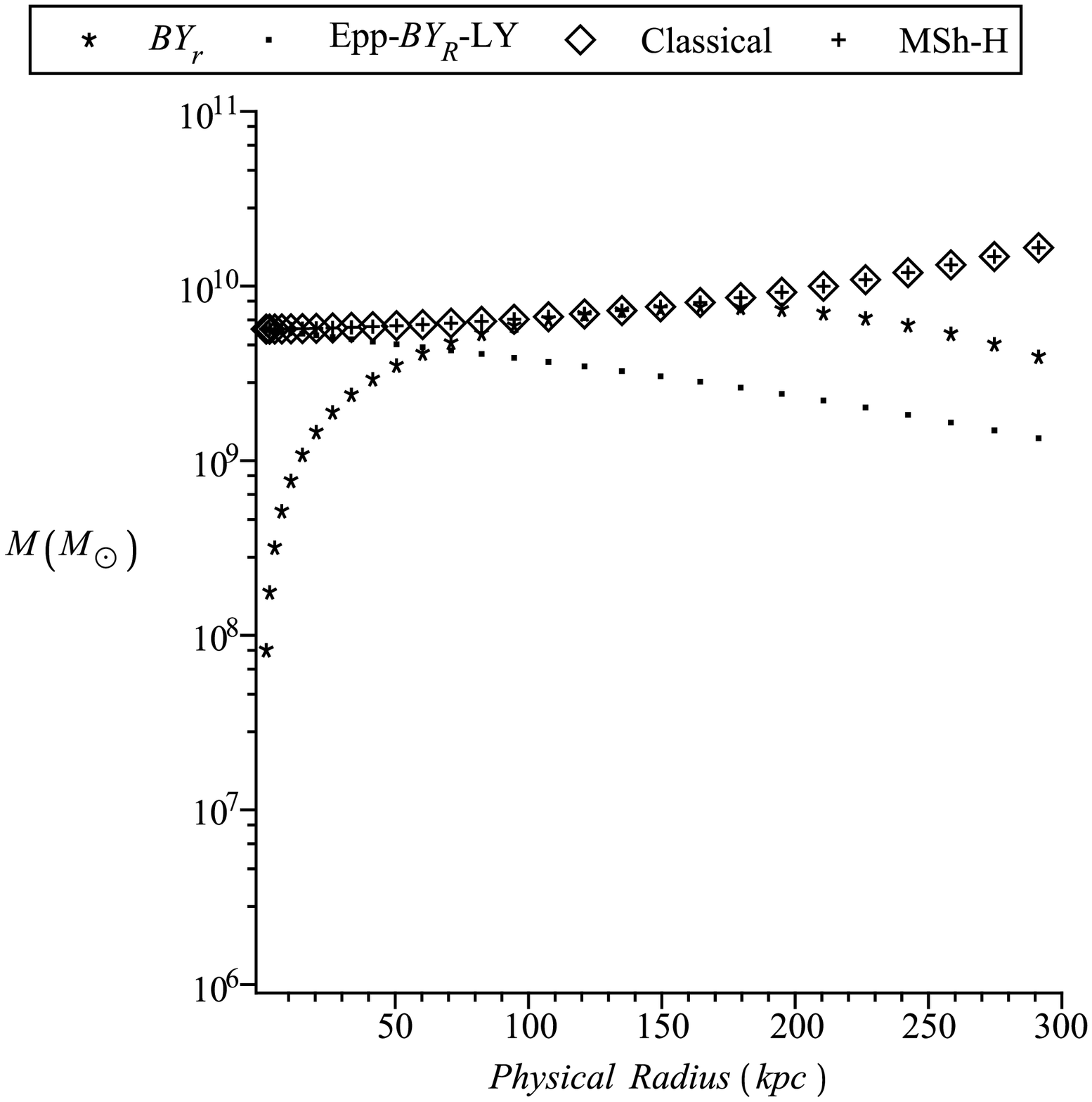}}\quad
\subfigure{\includegraphics[width=2in]{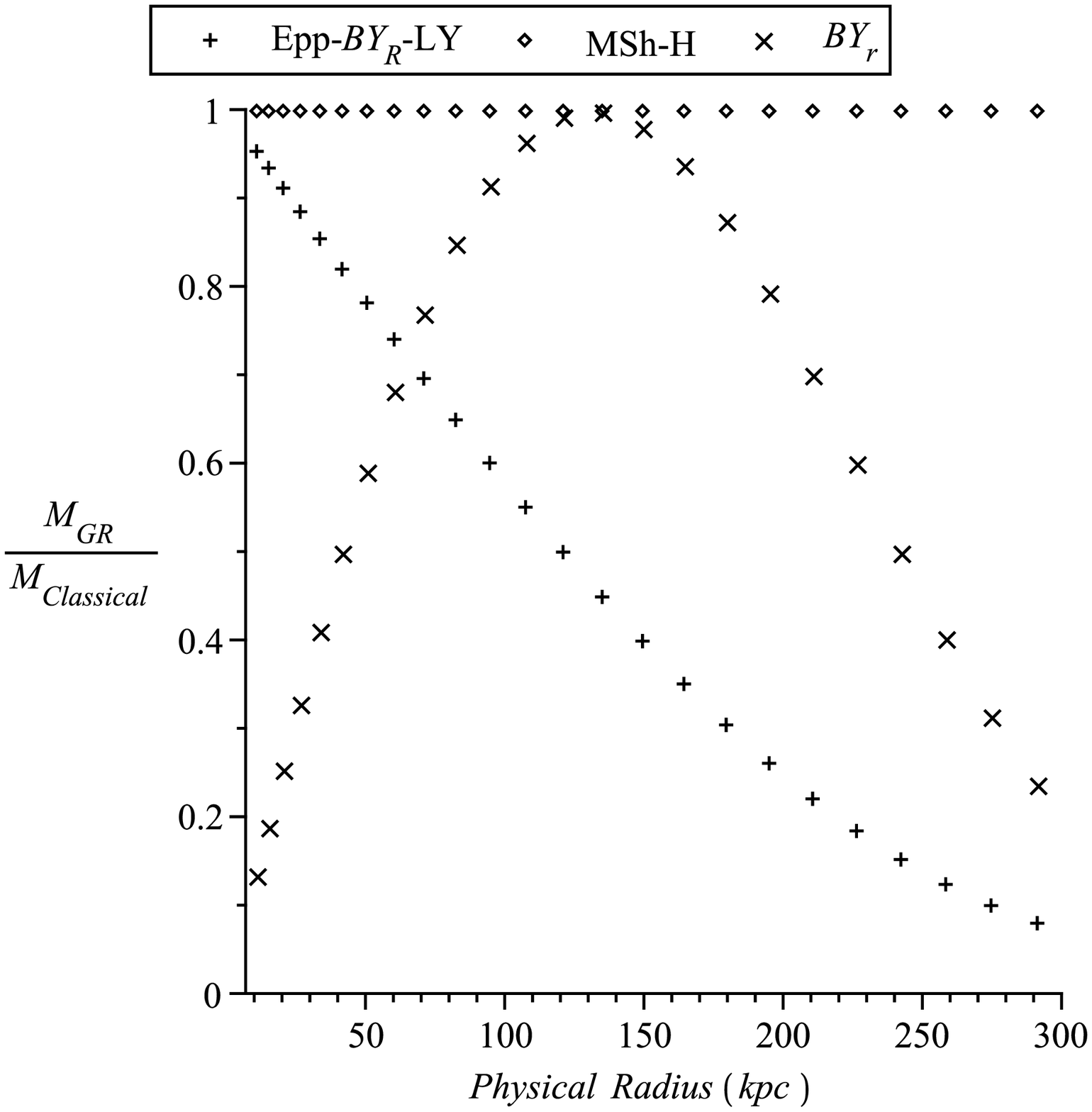} }}
\caption{  Quasi-local masses for a toy model of a galaxy corresponding to the density profile of the toy model } \label{fig3}
\end{figure}

\subsection{NFW model}

Now, we try a model structure with a NFW density profile \cite{nfw}. This density profile is used to find the areal
radius through an algorithm formulated in \cite{krasinki}, see also \cite{javad2}. To use this algorithm, the density
profile has to be specified at two different initial and final times, $t_i, t_f$ as a function of the coordinate
$r$. Now, the algorithm is based on the choice of $r$-coordinate such that $M(r) = r$. This is due to the fact
that $M(r)$ is an increasing function of $r$. Therefore, $E$ and $t_b$ become functions of $M$. The LTB functions $E(M)$
and $t_b(M)$ may then be extracted from the algorithm. For the initial time we choose the time of the last scattering surface:
$t_i\simeq 3.77\times10^5 yr$. The initial density profile should show a small over-density near the center imitating otherwise
a FRW universe. Therefore, we add a Gaussian peak to the FRW background density. We know already that having an over-density in
an otherwise homogeneous universe needs a void to compensate for the extra mass within the over-density region. Therefore, to
compensate this mass we subtract a wider gaussian peak. These density profiles may then be written as
\begin{equation}\label{yek}
\rho _{i}(r)=\rho _{crit}(t_{i})((\delta _{i}e^{-(\frac{r}{R_{i1}})^2}-b_{i})e^{-(\frac{r}{R_{i2}})^2}+{1}),
\end{equation}\
\begin{equation}\label{yek}
\rho _{f}(r)=\rho _{crit}(t_{f})((\frac{\delta _{c}}{(\frac{r}{r_{s}})(1+\frac{r}{r_{s}})^2}-b_{f})e^{-(\frac{r}{R_{f}})^2}+1),
\end{equation}\\
where $\delta_i$ is the density contrast of the Gaussian peak, $R_{i1}$ is the width of the Gaussian peak and $R_{i2}$ is the width of the
negative Gaussian profile at the initial time. The two constants $b_i$ and $b_f$ are then obtained by the mass compensation condition. The
results for the NFW density profile of a typical galaxy and the corresponding relativistic and Newtonian rotation curves are given
in Fig.(\ref{fig4}). We see again that both Newtonian and relativistic rotation curves coincide. \\
We have also calculated different general relativistic masses to see how different they are, although this does not change our
results as far as the rotation curve for a specific density profile is concerned. The results are shown in Fig.(\ref{fig5}). As in the
case of the toy model, we realize that the Misner-Sharp mass is almost identical to the Newtonian one. There is however substantial
difference to other general relativistic masses. \\

\begin{figure}[h]
\centering
\mbox{\subfigure{\includegraphics[width=2in]{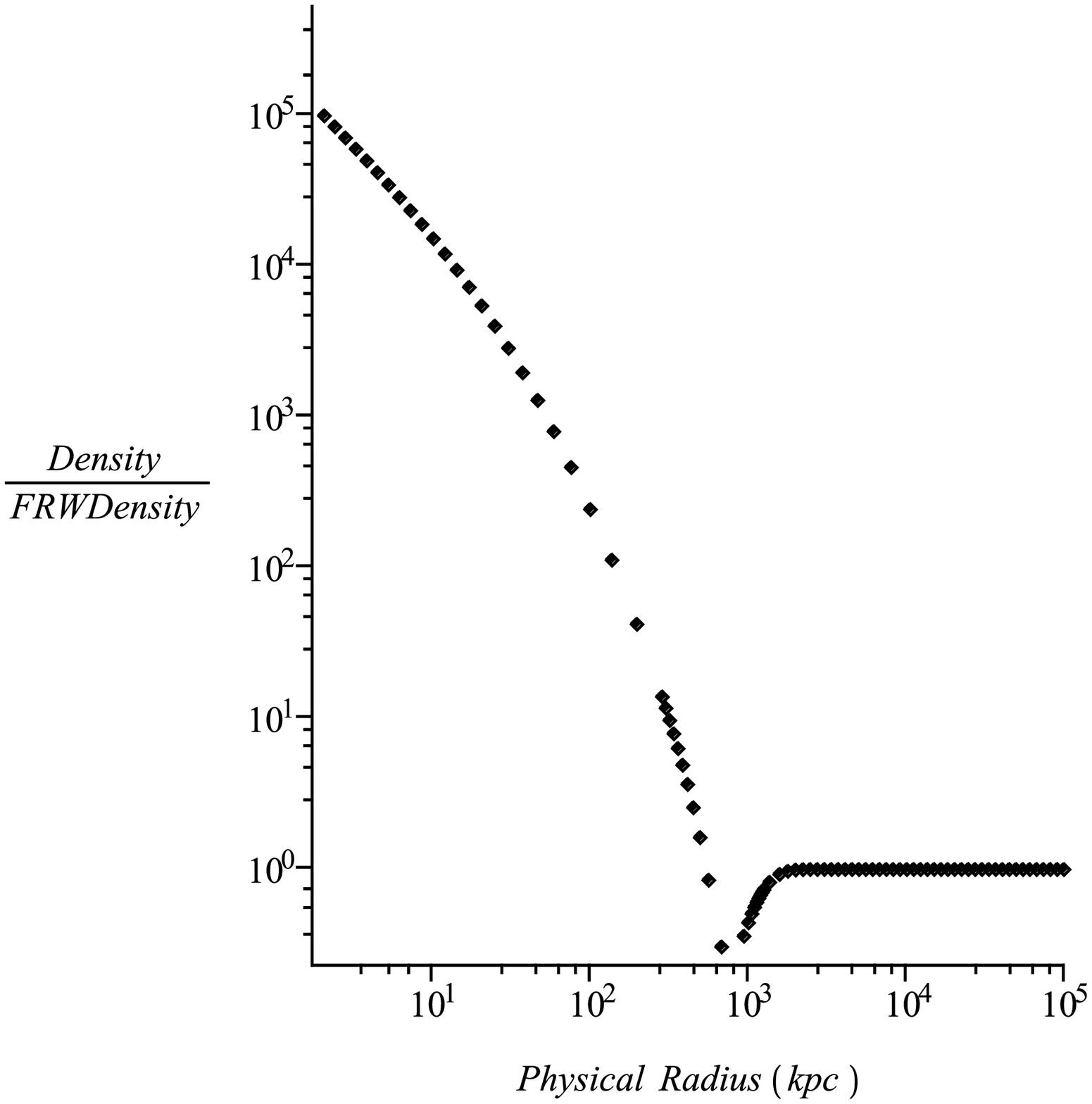}}\quad
\subfigure{\includegraphics[width=2in]{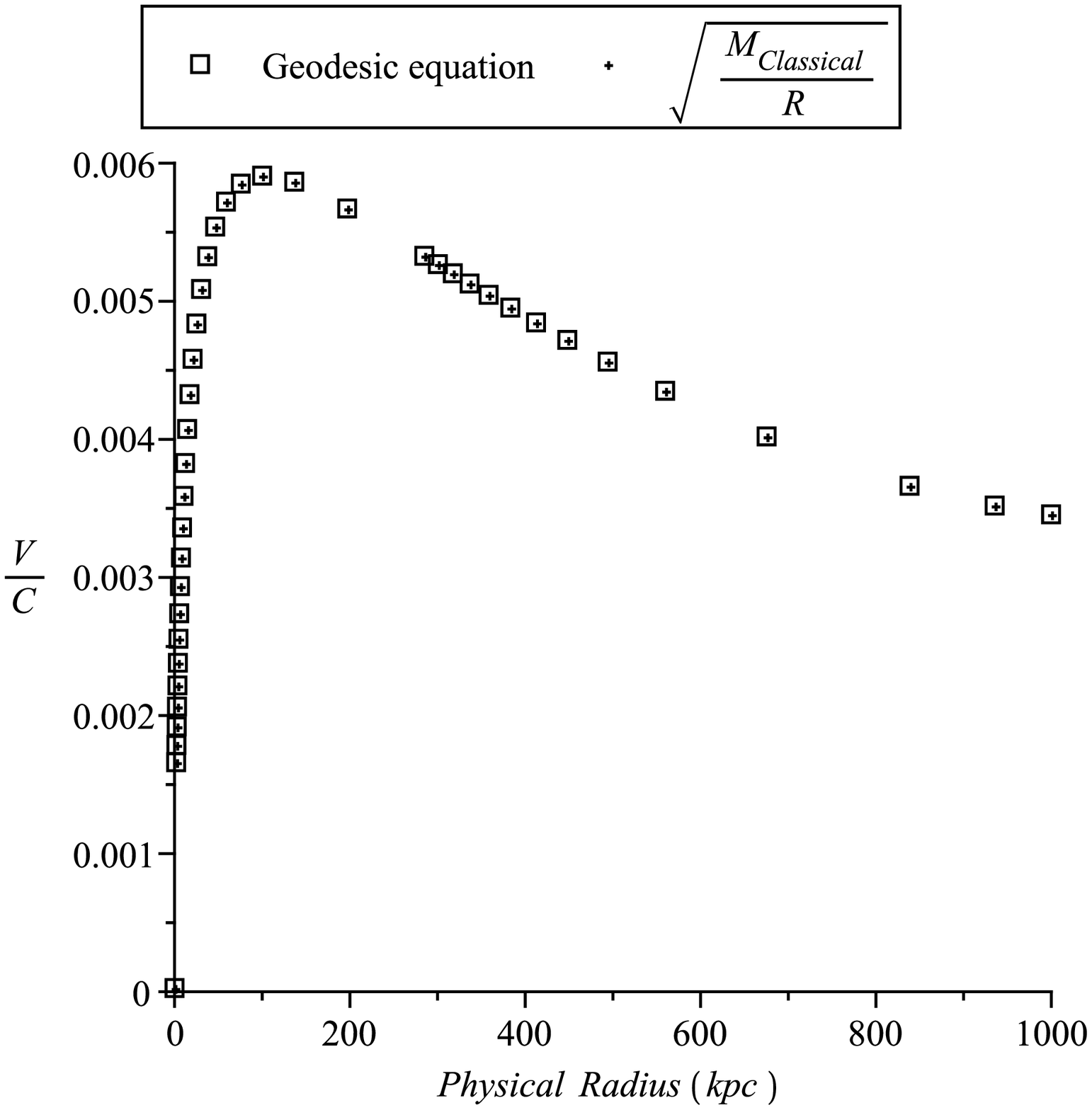} }}
\caption{Density profile for the NFW model of a galaxy and the corresponding Newtonian and relativistic rotation curve } \label{fig4}
\end{figure}

\begin{figure}[h]
\centering
\mbox{\subfigure{\includegraphics[width=2in]{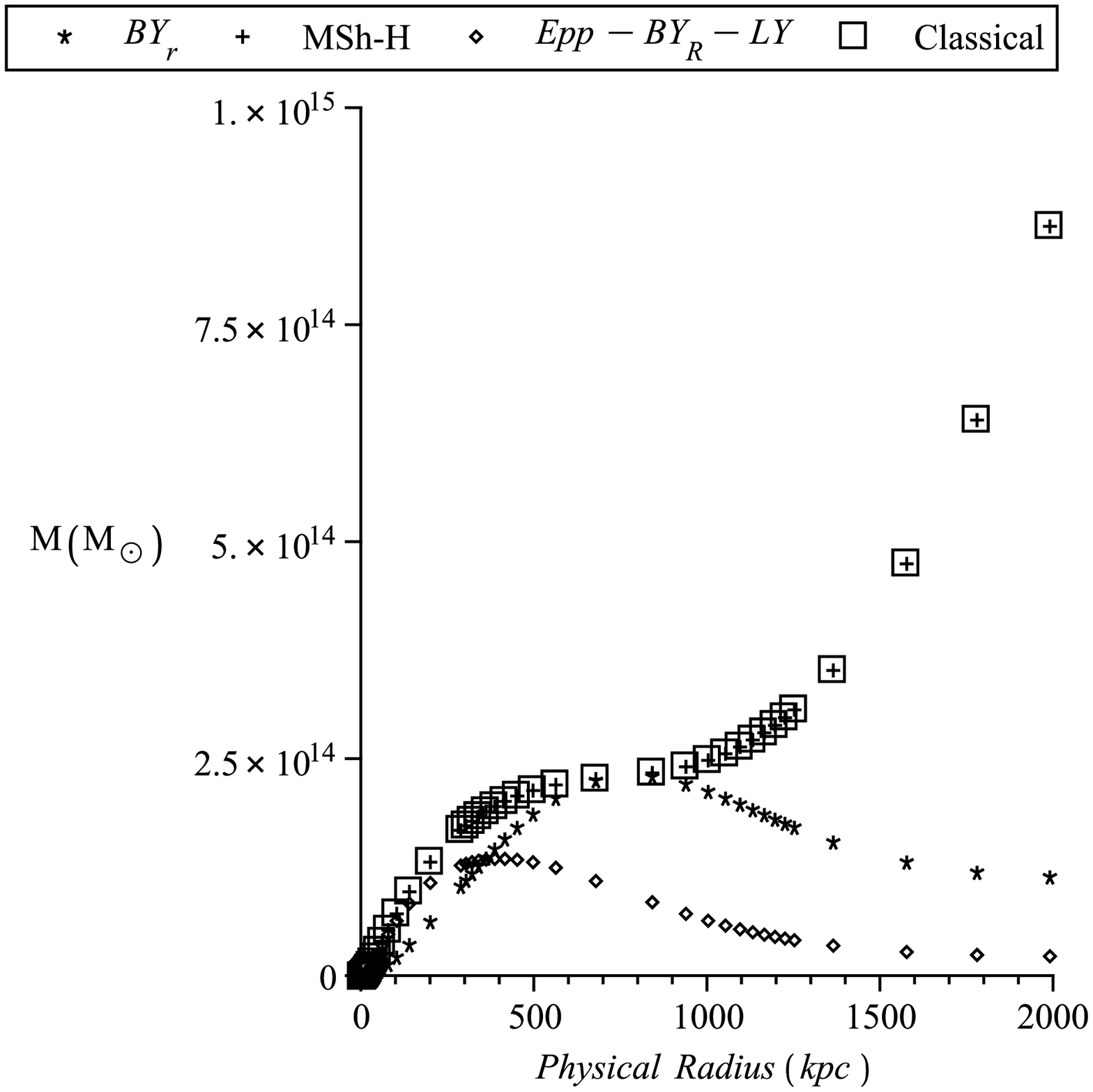}}\quad
\subfigure{\includegraphics[width=2in]{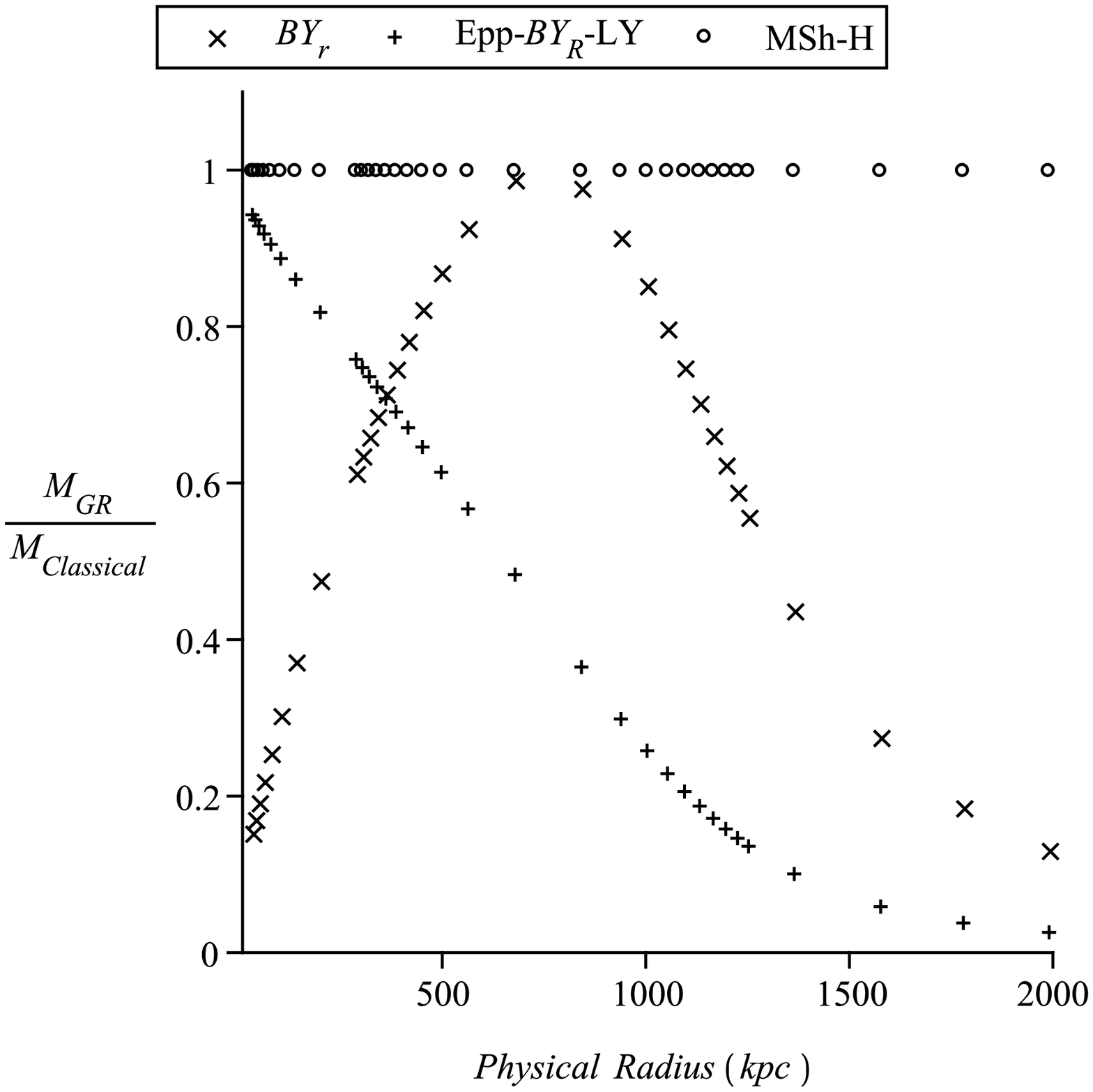} }}
\caption{ Ratio of GR masses to the Newtonian mass for the NFW density profile of a galaxy} \label{fig5}
\end{figure}

\section{discussion and conclusion }

We have tried for the first time to answer the question of how a relativistic rotation curve for a general relativistic structure within an otherwise expanding
universe looks like based on the Einstein equations. To achieve this, we have modeled our so-called cosmological structure using a LTB cosmological
solution tending to FRW universe at infinity and having an overdense structure at the center. After the original expansion, the overdense region starts
collapsing leading to a dense structure before going over to a FRW universe. The mass in-fall to the structure reduces at
late times leading to an almost static structure. This late time behavior gives us the possibility of defining quasi-circular orbits for particles
revolving around it. \\
We have then defined an algorithm how to obtain the relativistic rotation curve for a specific density profile, without relying on a mass definition
for the structure which is not unique in general relativity. The numerical procedure is tested first for a Schwarzschild metric where we have
shown that it is equivalent to the Newtonian rotation curve. We know already that in this case all the general
relativistic masses are equivalent to the Newtonian mass. In the case of a general relativistic cosmological structure we choose first a
toy model and then a NFW density profile. \\
Now, for our cosmological structure which is embedded in a dynamical FRW background, it turns out that the rotation curve at the galactic scale and
at far distances from the center of the structure (where the gravity is weak), is almost identical to the Newtonian one, which is the main result of our work.
We have also calculated the general relativistic mass definitions for our models to see how different they are, despite having a unique rotation
curve due to the specific density profile. \\

\section{ACKNOWLEDGMENT}

We would like to thank  Mojahed  Parsi Mood for helpful discussions and providing us his NFW model data.

\end{document}